\documentstyle[12pt]{article}
\include{epsf}
\epsfverbosetrue
\def\la{\langle }
\def\ra{ \rangle }

\newcommand{\beq}{\begin{equation}}
\newcommand{\eeq}{\end{equation}}
\newcommand{\bea}{\begin{eqnarray}}
\newcommand{\eea}{\end{eqnarray}}

\begin{document}
                                        \begin{titlepage}
\begin{flushright}
hep-ph/9906430
\end{flushright}
\vskip1.8cm
\begin{center}
{\LARGE False Vacuum Decay in QCD\\ 
\vskip0.6cm
within an Effective Lagrangian Approach
 }         
\vskip1.5cm
 {\Large T.~Fugleberg}
\vskip0.5cm
        Physics and Astronomy Department \\
        University of British Columbia \\
 6224 Agricultural Road, Vancouver, BC V6T 1Z1, Canada \\ 
        {\small e-mail:
fugle@physics.ubc.ca \\ }
\vskip1.0cm
{\Large Abstract:\\}
\end{center}
\parbox[t]{\textwidth}{
In an effective Lagrangian approach to QCD we nonperturbatively
calculate an analytic approximation to the decay rate of a false
vacuum per unit volume, $\Gamma/V$.  We do so for both zero and high
temperature theories.  This result is important for the study of the
early universe at around the time of the QCD phase transition.  It is
also important in order to determine the possibility of observing this
false vacuum decay at the Relativistic Heavy Ion Collider (RHIC).
Previously described dramatic signatures of the decay of false vacuum
bubbles would occur in our case as well.
}
\vspace{1.0cm}
\end{titlepage}
\section{Introduction}
Effective Lagrangian techniques have proven to be very valuable in
Quantum Field Theory.  There are two main types of effective
Lagrangian formulations currently in use.  The first type is the
Wilsonian effective action which describes the low energy dynamics of
the lightest particles in the theory.  The second type of effective
Lagrangian is defined as the Legendre transform of the generating
function of connected Green functions.  This formulation implements at
the Lagrangian level certain anomalous Ward identities relating vacuum
condensates of the fields and has been referred to as the anomalous
effective Lagrangian
\cite{QCDanalysis}.  This
type of approach is very useful for studying vacuum properties and was
first applied to supersymmetric (SUSY) gauge theories for both
Yang-Mills theory
\cite{VY} and full super-QCD\cite{TVY} and more recently to their 
non-supersymmetric 
versions \cite{1} \cite{QCD}.  The second type of effective Lagrangian
approach is the one used in this paper.

An effective Lagrangian, or more precisely the potential part, for
non-supersymmetric Yang-Mills theory was described in \cite{1} and for
non-supersymmetric QCD in \cite{QCD}.  This effective potential for QCD
for $N_f$ light flavors and $N_c$ colors was described in \cite{QCD}
and analyzed in detail in \cite{QCDanalysis}.  This study is interesting
for several reasons.  First, it provides a generalization of the large
$N_c$ Di Vecchia-Veneziano-Witten Effective Chiral Lagrangian
(ECL)\cite{Wit2} \footnote{This effective Lagrangian is of the second
type described above.} for arbitrary $N_c$ after integrating out the
heavy ``glueball'' fields.  Furthermore this approach to the
derivation of the VVW ECL fixes all dimensional parameters in terms of the
experimentally measurable quark and gluon condensates.  Second, this
effective Lagrangian approach makes it possible to address the problem
of $\theta$-dependence in QCD.  The problem of {$\theta$}-dependence
is also directly related to the problem of a realistic axion potential
since the axion comes from giving the $\theta$-parameter the status of
a dynamical field and the axion potential comes from the
$\theta$-dependence of the vacuum energy, $E_{vac}(\theta)$.  Third,
metastable vacua in the effective potential may provide a mechanism
for baryogenesis at the QCD scale\cite{BHZ}.  Finally, the previously
mentioned metastable vacua have dramatic signatures and would
potentially be observable at RHIC.  The final two reasons provided the
original motivation for the present work.

It is important to stress that the Anomalous Effective Potential is
simply a candidate for an effective potential for QCD.  We do
not wish to imply that effective chiral Lagrangians are in any
way wrong.  In particular, we are definitely not saying that the
Di Vecchia-Veneziano-Witten Effective Chiral Lagrangian is
wrong.  In fact one of the reasons for studying this theory is that
it agrees with the VVW ECL in the large $N_c$ limit.  This effective
Lagrangian is a generalization of the VVW to the case of finite $N_c$.

In the detailed study of the effective potential for QCD
\cite{QCDanalysis} it was found that the effective potential for the
phases of the chiral condensate exhibits cusp singularities as a
result of topological charge quantization\footnote{For specific values
of parameters of the theory which cannot be definitely determined from our
present knowledge}. These cusp singularities act as potential barriers
separating metastable vacuum states from the true physical
vacuum\footnote{For certain parameter values these metastable vacua
occur at every value of $\theta$.}.  The existence of these metastable
vacua leads to the well known phenomenon of false vacuum
decay\cite{Okun} which may have consequences in axion physics
\cite{axion}\cite{QCDanalysis}, baryogenesis\cite{QCDanalysis}\cite{BHZ} 
and/or many other unexplained phenomenon which may occur during/after
the QCD phase transition.  Domain wall solutions interpolating between
a metastable vacuum and the true vacuum were presented in
\cite{QCDanalysis} which will be used in this paper to calculate the
decay rate of the false vacuum.

These metastable vacua are somewhat controversial so some comment is
required.  First we would like to point out that nontrivial vacua have
been shown to exist in Yang Mills theories in the large $N_c$ limit
using the AdS/CFT correspondence\cite{Witten} and the same phenomenon
was observed\cite{Shifman2} in the analysis of soft breaking of
supersymmetric models.  Based on the original analysis of the VVW
ECL\cite{Wit2} and recent work\cite{Smilga} it is tempting to conclude
that metastable vacua only exist in QCD for $\theta\approx \pi$.  This
would contradict our claim that there are a series of metastable
vacua\footnote{The number of these metastable vacua goes to infinity
as $N_c$ goes to infinity.}  for $\theta=0$ and certain values of some
integer parameters of the theory.  However, as was stated above, the
effective Lagrangian we use is a generalization of the VVW ECL to
finite $N_c$ and it is not unreasonable to believe that this
generalization can lead to new features.  The effective potential we
use also represents a generalization of the effective potential of
\cite{Smilga} because of the inclusion of the $\eta^\prime$ and the
inclusion of two integer parameters, $p$ and $q$, about which there is
some controversy.  The first comment we would like to make is that if
we integrate out the singlet $\eta^\prime$ field from our form of the
effective potential we obtain the effective Lagrangian of
\cite{Smilga}. The second comment that we would like to make is that
if we choose $q=1$ and $p=N_c$, as some approaches to determining
their values would suggest, then there are no metastable vacua at
$\theta=0$ and all our results would agree with \cite{Smilga}.
However, the values $q=8$ and $p=11 N_c - 2 N_f$ arise in a number of
different approaches\cite{QCDanalysis}.  In this case we observe a
series of metastable vacua and cusps in the potentials for the chiral
fields even at $\theta=0$.  In any case it appears that the existence
of nontrivial vacua is a general phenomenon for gauge theories in the
strong coupling regime.  The main goal of this paper is to develop
techniques which could be useful in the study of this type of
phenomenon.

The decay rate of false vacua in large $N_c$ Yang Mills theory was
estimated in \cite{Shifman}.  Our calculation of the decay rate of a
false vacuum differs from \cite{Shifman} because it is valid for finite
$N_c$ andecause the heavy glueball degrees of freedom
have been integrated out in our approach,
while their calculation derives entirely from gluodynamics.  For more
comments on the elimination of heavy degrees of freedom see below. The
decay of a false vacuum was estimated in
\cite{Smilga} in the case that $\theta\approx\pi$ and $N_c$ is finite.
Our calculation differs from
\cite{Smilga} in the inclusion of the singlet $\eta^\prime$ field and
choice of parameters $q=8$ and $p=11 N_c - 2 N_f$.  As well, both of
these estimates only use the semiclassical approximation, while we
determine the effect of the first quantum corrections.  The form of
the semiclassical decay rate in our case is identical to that of
\cite{Shifman,Smilga}.  We make no numerical comparisons as the
false vacua involved are all different.

These false vacua and domain walls could lead to many interesting
consequences in the evolution of the early universe at around the time
of the QCD phase transition.  One example of this is related to
baryogenesis and dark matter, and is described in \cite{BHZ}.  The zero
temperature decay rate of the false vacua calculated in this paper
is relevant to this particular application.  As well, these metastable
states can hopefully be experimentally studied at RHIC and the
high temperature decay rate calculated in this paper would be
relevant to this research.  Bubbles of this false vacuum would display
CP odd signatures such as those described in \cite{Kharzeev} where the
large $N_c$ limit was assumed.

The purpose of the present paper is to determine how the decay rate
per unit volume, $\Gamma/V$, of the false vacuum depends on
physical parameters.  We will use several approximations along the way
to determine what we believe to be the dominant contribution within a
factor of about the order of unity.  It should be noted that while this is
an approximation it is nonperturbative
in the sense that it should contain contributions from all perturbative
diagrams.

Fermions (nucleons) could drastically change the results but since they
also drastically increase the difficulty of the calculation we will
leave them out in our first approximation.  The decay rate of the
metastable state could only be decreased by the inclusion of fermions
and thus our calculation is an estimate of the upper bound on the decay rate.

As well there is the consideration that intrinsic heavy degrees of
freedom might play a role.  It has been suggested\cite{KKS} that the
cusps in the potential are due to the integration out of heavy degrees
of freedom and that the presence of cusps invalidates the construction
of the domain wall from the effective Lagrangian.  This would mean
that the heavy degrees of freedom must be included for the complete
calculation.  However, as above, the decay rate can only be decreased
by the inclusion of these heavy degrees of freedom and we ignore their
effects in our estimate of the upper bound on the decay rate.

Even without consideration of these interesting applications, our
method of calculation of the determinantal prefactor is useful as an
alternative to previous methods.  Previous calculations
\cite{Cottingham} \cite{Kiselev} \cite{Baacke} \cite{Strumia}
of bubble nucleation rates use a particular method for calculating the
determinant ratio of operators of the form:
\beq
{\cal M}= -\nabla^2 + \omega^2 + \alpha V(r),
\eeq 
involving a theorem from \cite{Cottingham}.  We prefer to use a more direct approach. 
Our procedure provides a method for
obtaining both analytical approximations and exact numerical
calculations for this determinantal prefactor.
Our numerical approximation involves only numerical integration unlike the
method of \cite{Surig} which also involves numerical solution of 
differential equations and might not be reliable in some cases
involving non-smooth perturbation potentials.
Our methods are more along the lines of \cite{CLMW} but are sufficiently
different to constitute independent results.
As well we have used the same methods to calculate the decay rate in the
zero temperature theory which has not been done before. 

The paper is organized as follows.  In Sect.2 we review the structure
of the effective potential and the domain wall
solution\cite{QCDanalysis}.  In Sect.3 we determine the semiclassical
approximation to the decay rate for both zero and high temperature.
We evaluate the first quantum corrections at zero temperature in
Sect.4 and at high temperature in Sect.5.  Finally in Sect.6 we
discuss the implications of our results for baryogenesis and dark
matter and for observations of parity odd bubbles at RHIC.

\section{The Effective Potential}
The effective potential\cite{QCD}\cite{QCDanalysis} for QCD is:
\bea
\label{8}
W_{eff}(U,U^{+}) =  - \lim_{V \rightarrow 
\infty} \; \frac{1}{V} \log \left\{ 
 \sum_{l=0}^{p-1} \exp \left[ 
V E \cos \left[ - \frac{q}{p} ( \theta - i \log  \det \, U ) 
+ \frac{2 \pi}{p}
\, l \right]  \right. \right. \nonumber \\
\left. \left. + 
\frac{1}{2} V \, Tr ( M U + M^{+} U^{+} ) \right] \right\}   
\; ,
\eea
where the light\footnote{Note that the
$\eta^\prime$ is not really very light, but it enters the theory in this
way.} matter fields are 
described by the unitary matrix, $U_{ij}$, corresponding to the
phases of the chiral condensate:
\begin{equation}
 \langle\bar{\Psi}_{L}^{i} 
\Psi_{R}^{j} \ra 
=  - | \langle\bar{\Psi}_{L} \Psi_{R} \ra | \, U_{ij},
\end{equation} 
with\footnote{Note that mixing of the flavor eigenstates is ignored
at this level.}:
\beq
\label{BranchCondition}
U = \exp \left[ i \sqrt{2} \, \frac{\pi^{a} \lambda^{a} }{f_{\pi}} + i
\frac{ 2}{ \sqrt{N_{f}} } \frac{ \eta'}{ f_{\eta'}} \right]
\; \; , \; \; 
U U^{+} = 1 \; .
\eeq
where $ \lambda^a $ are the Gell-Mann matrices of $SU(N_f)$, $ \pi^a $
is the octet of pseudoscalar fields (pions, kaons and the eta meson) and
$\eta^\prime$ is the $SU(N_f)$ singlet pseudoscalar field.
$ M = diag (m_{i}  | \la \bar{\Psi}^{i} \Psi^{i} 
\ra | )$, V is the 4-volume and the integers p and q are relatively prime
parameters.  The values of the parameters p and q are not known as
different proposals for their determination lead to different
values\cite{QCDanalysis}.  We will not use specific values of p and q
aside from the restriction that $q\neq 1$.  We would like to mention
that the values $p=11 N_c - 2 N_f$ and $q=8$ arise in a number of
different approaches and that $q/p \sim 1/ N_c$ for the U(1) problem
to be solved.  $E=\langle b\alpha_s/(32\pi) G^2 \rangle$ where
$b=\frac{11}{3}N_c -\frac{2}{3}N_f$ is the Gell-Mann - Low 
$\beta$-function of QCD.  The physical input to this equation are the
values of the vacuum condensates, $\langle \alpha_s/\pi\: G^2
\rangle=0.012 \:\mbox{GeV}^4$ and $\langle \bar{\Psi} \Psi \rangle=
(240 \: \mbox{MeV})^3$, and  the quark masses.  We use the values $ f_{\pi}
= 132 \; MeV $ and $ f_{\eta^\prime} = 86 \; MeV $.

We will take this potential as our starting point motivated by four of its
most important properties (for details see \cite{QCDanalysis}):

i) it correctly reproduces the VVW effective Chiral Lagrangian
\cite{Wit2} in the large $N_c$ limit.

ii)it reproduces the anomalous conformal and chiral Ward identities of QCD.

iii)it reproduces the known $\theta$-dependence for small $\theta$-angles
\cite{Wit2} but leads to $2\pi$ periodicity in $\theta$ of physical observables. 

iv) the related effective Lagrangian for pure gluodynamics\cite{1} has
the nice property:
\beq
\left. \frac{d^{2k} E_{vac}(\theta)}{d\theta^{2k}} \right| \sim (1/N_c)^{2k},
\eeq
which was advocated earlier by Veneziano for the $U(1)$ problem 
to be resolved \cite{Veneziano}.

This effective potential is not representable by a single analytic
function in the $V\rightarrow\infty$ limit.  The thermodynamic limit
selects, at each value of $\theta - i\log \det U$, one particular
branch ({\it ie.} a particular value of the integer $l$) and cusp
singularities occur where the branches coincide.

The effective potential for the chiral phases of the matter fields,
becomes:
\beq
\label{13}
W_{eff}^{(l)} = - E \cos \left( - \frac{q}{p} \theta + 
\frac{q}{p} \sum \phi_{i} + \frac{2 \pi}{p} \, l \right) 
- \sum M_{i} \cos \phi_i  \; \; , \; \;   l= 0,1, \ldots , p-1,
\eeq
if 
\beq
\label{14}
(2 l - 1) \frac{\pi}{q} \leq \theta - \sum \phi_i < (2 l + 1) 
 \frac{\pi}{q} \; \;  .
\eeq
when we take, $U=diag(\exp i\phi_q)$~\footnote{This is not a
restriction since the quark mass matrix can always be diagonalized.}.
This is now a piecewise smooth potential for the phases of the chiral condensate
with cusp singularities.

At this point some comment is required about the removal of the singlet
$\eta^\prime$ field from the effective potential. If we minimize the
potential for $l=0$ with respect to the singlet field, $\sum \phi_i$, we clearly
obtain the solution:
\begin{equation}
\sum \phi_i=\theta.
\end{equation}
Integrating out the singlet field amounts to substituting this solution
in the effective potential leading to:
\beq
\label{13b}
W_{eff}(\phi_1,\phi_2) = 
-  M_{1} \cos \phi_1 -  M_{2} \cos \phi_2 - M_{3} \cos (\theta-\phi_1-\phi_2)
\eeq
which upon the substitutions, $\phi_{1} \rightarrow \theta/3-\alpha$,
$\phi_{2} \rightarrow \theta/3-\beta$ and $M_i \rightarrow m$ becomes
exactly the potential given in Equation (2) of \cite{Smilga}.
Therefore, as we asserted in the Introduction, integrating the singlet
field out of our effective potential we obtain the same effective
potential as in \cite{Smilga}.

For $q\neq 1$ there exist metastable vacuum states in addition to the
lowest energy physical vacuum which leads to false vacuum decay.  For
our purposes we will use the simplified setting where $\theta=0$ and
$N_f=3$, equal quark masses $m_i\equiv 4\:\mbox{MeV}$~\footnote{Note
that for $q\neq 1$ there is no phenomenological sensitivity to the
values of the light quark masses and we are allowed to choose the
quark masses to be equal. For details see \cite{QCDanalysis}.}, and equal chiral phases $\phi_i=\phi$.
This amounts to studying only radial motion in the $\phi$-space.  We
analyze the problem in the spirit of Ref.\cite{Okun} and only consider
transitions between the lowest energy metastable state and the
physical vacuum.  The results should be easily generalizable to other
transitions.

\begin{figure}
\epsfysize=3in
\epsfbox[ -40 416 604 772]{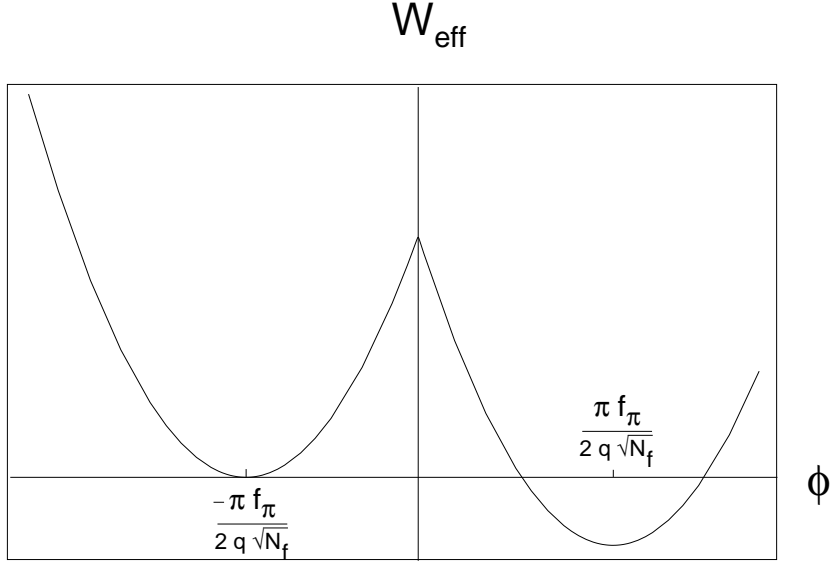}
\caption{Effective potential at $ \theta = 0 $ for equal chiral phases 
  $\phi=\phi_u=\phi_d=\phi_s$.}
\label{tunneling}
\end{figure}

\begin{figure}
\epsfysize=3in
\epsfbox[-50 463 556 782]{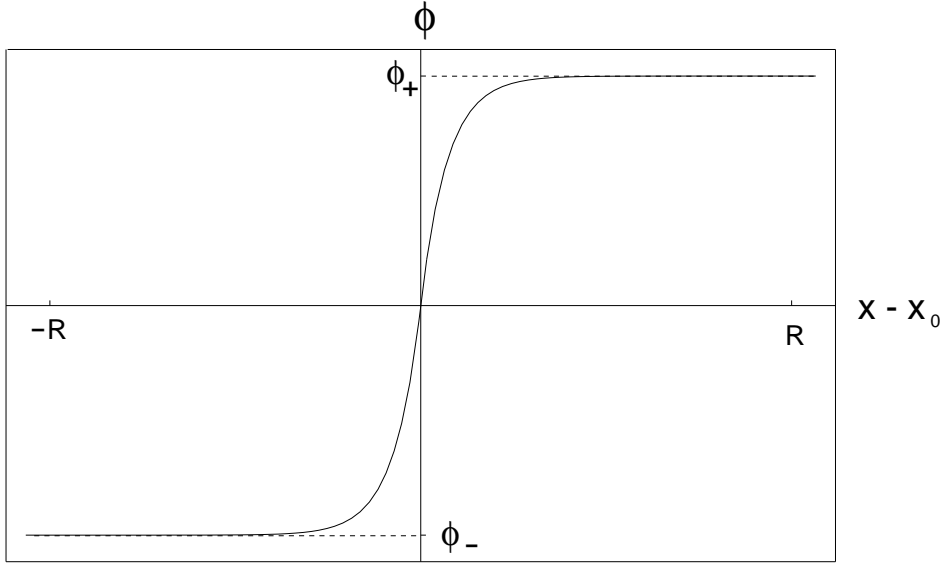}
\caption{Domain wall profile.} 
\label{tunneling_solution}
\end{figure}

For ease of calculation we rescale and shift the chiral field $ \phi
\rightarrow (2 / f_{\pi} \sqrt{N_f} ) \phi -
\pi/(q N_f ) $ in 
order to have the standard normalization of the kinetic term
and a symmetrized form of the potential.
The effective potential for $ \theta = 0 $
becomes
\bea
\label{18a}
W (\phi) = \left\{ \begin{array}{ll}
E \left[ 1 - \cos \left( \frac{ 2 q \sqrt{N_f}}{p f_{\pi}} \phi 
- \frac{ \pi}{p} \right) \right]
-  M f(\phi)  
  & \mbox{if $ \phi \geq 0   $ } \\ \nonumber 
E \left[ 1 - \cos \left( \frac{ 2 q \sqrt{N_f}}{p f_{\pi}} \phi
+ \frac{ \pi}{p} \right)  \right]
-  M f(\phi) 
& \mbox{if $ \phi \leq  0   $ } 
\end{array} 
\right., \nonumber \\
f(\phi) =  N_f \left[ \cos \left( \frac{2}{f_{\pi} \sqrt{N_f}} \phi - 
 \frac{ \pi}{q N_f } \right) - \cos \left( \frac{2 \pi}{q N_f} 
\right) \right].
\eea
The effective potential (\ref{18a}) has a global minimum
at $ \phi_+ =  (\pi f_{\pi})/(2 q 
\sqrt{N_f} ) $ and a local minimum at $ \phi_- = - (\pi f_{\pi})/(2 q 
\sqrt{N_f} ) $, with a cusp singularity between them, (see 
Fig. \ref{tunneling}).
The minima are interpreted as two vacua separated by a high potential
barrier ($\sim G^2$) which is fairly wide, while the energy splitting,
$\Delta E$ between the states is fairly small in comparison,
\beq
\label{200}
 \Delta E =   m_q N_f \left| \la \bar{\Psi}\Psi\ra \right| 
\left( 1-\cos  \frac{2\pi}{qN_{f}} \right)  +0(m_q^2).
\eeq 
Therefore we can use the thin wall approximation \cite{Okun} in our
calculations.  The domain wall solution in this approximation
corresponding to the effective potential (\ref{18a}) is:
\bea
\label{19a}
\phi_{d.w.} (x) &=& \frac{p f_{\pi} }{2 q \sqrt{N_f}} \left[ - \frac{\pi}{
p} + 4 \arctan \left\{ \tan \left( \frac{\pi}{4 p} \right) \exp [
\mu (x - x_0)] \right\}  \right] \; \; \; \; \mbox{if } x < x_0  
\nonumber \\
&=& \frac{p f_{\pi}}{2 q \sqrt{N_f}} \left[  \frac{\pi}{
p} - 4 \arctan \left\{ \tan \left( \frac{\pi}{4 p} \right) 
\exp \left[
- \mu (x - x_0) \right] \right\}  \right] \; \; \; \; \mbox{if } x > x_0 \; ,  
\eea 
where $ x_0 $ is the position of the center of the domain wall and 
\beq
\label{19c}
\mu \equiv\left. \sqrt{\frac{d^2 W_{eff}}{d\phi^2}}\right|_{min} = 
\frac{2 q \sqrt{N_f} \sqrt{E}}{ p f_{\pi} }  
\eeq 
is the inverse width of the wall.  The solution (\ref{19a}) is shown
as a function of $ x - x_0 $ in Fig. (\ref{tunneling_solution}).
Its first derivative is continuous at $ x = x_0 $, but the 
second derivative exhibits a jump.
The wall surface tension is given by:
\beq
\label{19d}
\sigma=\frac{4 p}{q \sqrt{N_f} }  
f_{\pi}\sqrt{\la \frac{ b \alpha_s }{ 32 \pi} G^2 \ra }
\, \left( 1 - \cos \frac{\pi}{ 2 p} \right) +0(m_q f_{\pi}^2 ).
\eeq 

In what follows we use this domain wall solution in our calculation
of the decay rate per unit volume of the false vacuum for both zero
and high temperature field theory.  The next section describes the
semiclassical calculation.

\section{Semiclassical Theory}

The fate of the false vacuum was discussed in \cite{Okun} and the expression of
$\Gamma/V$ has the form:
\beq
\Gamma/V=A e^{-B/\hbar}[1+O(\hbar)].
\label{semiclass}
\eeq

The semiclassical approximation at zero temperature tells us that B is
given by the Euclidean action of $\phi$:
\beq
B=S_4=\int d\tau d^3\vec{x}\:\left[\frac{1}{2}(\frac{\partial\phi}
{\partial\tau})^2+
\frac{1}{2}(\nabla\phi)^2+U(\phi)\right],
\eeq
where $U(\phi)$ is the potential, $W_{eff}$, for the chiral phases
described in the previous section neglecting the energy difference,
$\Delta E$, between the two vacua:
\bea
U(\phi) = \left\{ \begin{array}{ll}
E \left[ 1 - \cos \left( \frac{ 2 q \sqrt{N_f}}{p f_{\pi}} \phi 
- \frac{ \pi}{p} \right) \right]  
  & \mbox{if $ \phi \geq 0   $ } \\ 
E \left[ 1 - \cos \left( \frac{ 2 q \sqrt{N_f}}{p f_{\pi}} \phi
+ \frac{ \pi}{p} \right)  \right]  
& \mbox{if $ \phi \leq  0   $. } 
\label{symmetricpotential}
\end{array} 
\right.
\eea
In order for this to be finite we must have
$\lim_{r\rightarrow\infty}\phi(r)=\phi_-\equiv\pi f_\pi/2q
\sqrt{N_f}$ where $r=\sqrt{x^2+y^2+z^2+t^2}$.  The solution of this problem
 is the four
dimensional equivalent to the solution which Coleman calls ``the
bounce"\cite{Okun}.  It describes a bubble of the true vacuum in the
false vacuum at the origin which forms, grows to a maximum size and
then shrinks to nothing again ({\it ie.} an O(4) invariant bubble). In the
thin wall approximation the bounce solution is:
\beq
\phi_b(r)=\left\{
  \begin{array}{l}
    \phi_- \: \: \mbox{for} \: \: r \ll R \nonumber \\
    \phi_{d.w.}(r-R) \: \: \mbox{for} \: \: r \approx R \\
    \phi_+ \: \: \mbox{for} \: \: r \gg R \nonumber
  \end{array} \right. ,
\eeq
and the value of B is calculated to obtain\cite{Okun}:
\beq
\label{19f}
\Gamma/V\propto \exp \left( - S_4[\phi_b] \right)=\exp \left( 
-\frac{27\pi^2\sigma^4}{2(\Delta E)^3} 
\right),
\eeq
where $\sigma$ is the domain wall energy and:
\beq
\label{19g}
S_4[\phi_b] = 
\frac{ 3^3 \cdot 2^7 \cdot \pi^2 p^4}{q^4 N_{f}^5} \, \frac{f_{\pi}^4
E^2}{ M^3} \, \frac{ \left( 1 - \cos \frac{\pi}{2p} \right)^4 }{
\left( 1 - \cos \frac{ 2\pi}{q N_f} \right)^3 } \simeq 
\frac{27}{256} \, \frac{\pi^4 q^2 N_f}{p^4} \, \frac{
f_{\pi}^4  \la \frac{ b \alpha_s }{ 32 \pi} G^2 \ra^2}{ m_{q}^3 
\left| \la \bar{\Psi}\Psi\ra \right|^3 } \; 
\eeq
is the Euclidean action of the 4D bounce solution.

The thin wall approximation is valid because the radius of the 4D bubble,
which is found to be $R=3 \sigma /\Delta E$ by minimizing the value 
$S_4[\phi_b]$ , is much larger than the width of 
the domain wall, $1/\mu$.

For finite temperature QCD the semiclassical approximation is slightly
different.  At sufficiently high temperature the bubble solution becomes 
a stable O(3) invariant bubble with radius\cite{Linde2}:
\beq
R(T)=\frac{2S_1(T)}{\Delta E}.
\eeq
where the temperature dependence of the domain wall energy, $S_1(T)$,
is not known.  In this case the calculation of B gives\cite{Linde2}:
\beq
\Gamma/V\propto \exp ( - S_3[\phi_b]/ T )=\exp \left( -\frac{16\pi 
S_1(T)^3}
{3(\Delta E)^2 T}
\right) .
\label{HighTSemiclassical}
\eeq 
It should be noted that this decay rate is for the ground state of the
metastable well.  The decay rate for excited energy states above
the metastable vacuum via thermally activated transitions while similar in form
is not the same\cite{Affleck}.

This completes the semiclassical analysis of the decay rate.  In the next section
we calculate the quantum corrections to the decay rate in the zero temperature
theory.

\section{Quantum Corrections at Zero Temperature}
The quantum corrections at zero temperature correspond to the coefficient A
in Eq.(\ref{semiclass})\cite{Callan}:
\beq
A=\left|\frac{\det[-\partial_\mu \partial^\mu+U''(\phi_b)]}{\det[-\partial_\mu 
\partial^\mu+U''(\phi_-)]}
\right|^{-1/2}.
\eeq
The spectrum of the operator in the numerator consists of a discrete
spectrum with zero and negative eigenvalues and a continuous positive
eigenvalue spectrum starting at $\omega^2\equiv U''(\phi_-)$.  These
two parts of the spectrum must be analyzed separately and it can be
shown that this factor separates into three parts:
\beq
A=\left(\omega^4 \int d^4x \left( \frac{B}{2\pi}\right)^{2} \right)
\frac{\omega} {\sqrt{\lambda_-}}
\left|\frac{\det'[-\partial_\mu \partial^\mu+U''(\phi_b)]}{(\omega^{-2})^5\det[-\partial_\mu 
\partial^\mu+\omega^2]}
\right|^{-1/2} .
\label{A}
\eeq
The first term comes from the zero eigenvalues.  The second term comes
from the negative eigenvalue.  The third term is the determinant of
the continuous positive eigenvalue spectrum where $\det'$ means that
the zero and negative eigenvalues are to be omitted. With these
eigenvalues omitted the perturbed operator in the numerator has five
less eigenvalues in the spectrum because five of the eigenvalues of
the unperturbed operator in the denominator have become part of the
discrete spectrum of the perturbed operator.  Assuming these
eigenvalues have originated from the bottom of the unperturbed
continuous spectrum we divide through the third term by a factor of
$\omega^2$ for each omitted eigenvalue.  The contributions from the
zero and negative eigenvalues are normalized with a factor of $\omega$ keeping
each of the three terms dimensionless.

\subsection{Positive Eigenvalues}
The contribution of the positive eigenvalues requires the evaluation
of the determinant ratio:
\beq
 \left|\frac{\det'[-\partial_\mu \partial^\mu+U''(\phi_b)]}{(\omega^{-2})^5\det[-\partial_\mu
 \partial^\mu+\omega^2]} \right|.
\label{ratio}
\eeq
However, since the $(\omega^{-2})^5$ in the denominator corresponds to 
a set of measure zero in the continuous eigenvalue spectrum we can omit
it and the notation det' which indicates omission of a discrete set
of eigenvalues.

\begin{figure}
\epsfysize=1in
\epsfbox[-50 671 327 743]{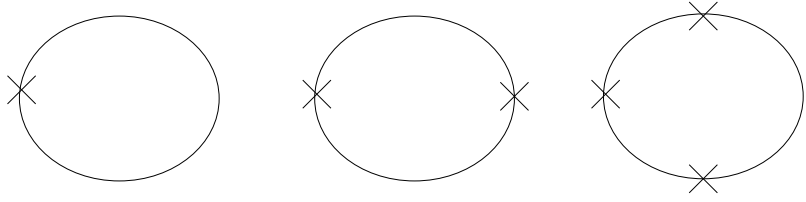}
\caption{Expansion of the partition function.} 
\label{expansionfigure}
\end{figure}

This determinant ratio will turn out to be infinite and to
obtain a finite answer we must divide by an infinite factor.  The
determinant in the numerator can be expanded in the following way:
\bea
&&\det|-\partial_\mu \partial^\mu+U''(\phi_b)|= \exp\left[ Tr \log 
\left\{
-\partial_\mu \partial^\mu + \omega^2 + V_{pert}(r) \right\} \right]
\label{loopexpansion}
\\
&&= \det|-\partial_\mu \partial^\mu+\omega^2| \times \nonumber \\
&&\exp \: Tr \left[
\frac{V_{pert}(r)}{-\partial_\mu \partial^\mu+\omega^2}
-\frac{1}{2}\left(\frac{V_{pert}(r)}{-\partial_\mu \partial^\mu+\omega^2}
\right)^2+
\frac{1}{3}\left(\frac{V_{pert}(r)}{-\partial_\mu \partial^\mu+\omega^2}
\right)^3 +
\cdots \right]. \nonumber
\eea
It should be noted that the determinant is equal to the partition
function of the self interacting massive scalar particle and that the
second factor in the last line is expanded up to one loop
contributions with three interactions with the effective external
potential(see Fig. \ref{expansionfigure}).  Tracing over a Cartesian
basis we can see that the one and two interaction contributions are
divergent but the three interaction contribution is finite:
\newpage
\bea
Tr \left[ \frac{V_{pert}(r)}{-\partial_\mu \partial^\mu+\omega^2}
\right]&=& \int d^4x \;d^4k \; \frac{V_{pert}(r)}{k^2+\omega^2}
\label{1loop},\\ 
Tr \left[ \left(\frac{V_{pert}(r)}{-\partial_\mu
\partial^\mu+\omega^2} \right)^2\right]&=& \int \;d^4k \;d^4p \;
\frac{{\cal V}_{pert}(k)}{(k^2+\omega^2)} \frac{{\cal
V}_{pert}(-k)}{((k+p)^2+\omega^2)}\label{2loop},\\
 Tr \left[
\left(\frac{V_{pert}(r)}{-\partial_\mu \partial^\mu+\omega^2}
\right)^3\right]&=& \int \;d^4k \;d^4p \;d^4q \;
\frac{{\cal V}_{pert}(k)}{(k^2+\omega^2)} 
\frac{{\cal V}_{pert}(p)}{(p^2+\omega^2)} 
\frac{{\cal V}_{pert}(-k-p)}{((k+p+q)^2+\omega^2)}.
\label{3loop}
\eea
where ${\cal V}_{pert}(k)$ is the Fourier transform of the
$V_{pert}(r)$. We will only get a finite answer if we divide through
by the infinite factors.  

The actual calculation is most easily done
using hyperspherical coordinates in four dimensions, $(r,\theta,\phi
,\psi)$, and expanding the eigenfunctions $\chi(r,\theta,\phi,\psi)$
in terms of the 4D hyperspherical harmonics:
\beq
\chi(r,\theta,\phi,\psi)=
\sum_{n=0}^\infty \sum_{l=-n}^{n} \sum_{m=-|l|}^{|l|} C_{nlm}\frac{u(r)}
{r^{3/2}} 
Y_{nlm}(\theta,\phi,\psi),
\label{expansion}
\eeq
which are discussed in the appendix.

In this situation the Laplacian, when operating on each term of
(\ref{expansion}) with quantum number `n', becomes:
\beq
\partial_\mu \partial^\mu  \rightarrow \frac{1}{r^{3/2}} \left[ \frac{d^2}
{dr^2}- 
\frac{4n^2+8n+3}{4r^2} \right]r^{3/2}\equiv \frac{1}{r^{3/2}} {\cal D}_n 
r^{3/2}.
\label{laplacian}
\eeq
There are 2n(n+2)+1 such terms corresponding to different values of
`l' and `m' but with the same eigenvalue.  Therefore:
\beq
\det[-\partial_\mu \partial^\mu+U'']=\prod_{n=0}^\infty 
(\det[-{\cal D}_n+U''])^{2n(n+2)+1}.
\label{nprod}
\eeq

First consider the denominator of (\ref{ratio}), $\det[-{\cal D}_n+\omega^2]$.
In order to calculate this
determinant we need to solve the eigenvalue equation:
\beq
[{\cal D}_n-\omega^2 + \lambda]u(r)=0.
\eeq
The solutions to this differential equation that are well behaved at 
$r=0$ and $r=\infty$ are:
\beq
u(r)= (\omega_\lambda r)^{3/2} \jmath_n(\omega_\lambda r)\equiv
\sqrt{\frac{\pi\omega_\lambda r}{2}} J_{n+1}(\omega_\lambda r),
\eeq
where $\omega_\lambda\equiv 
\sqrt{\lambda-\omega^2}$.  $\jmath_n$ are the 4D analogs of the spherical 
Bessel functions and $J_{n+1}(x)$ are Bessel functions of the first 
kind.
These solutions become purely oscillatory as $r \rightarrow \infty$.
Notice that this solution is only well defined for $\lambda>\omega^2$
and indeed there are no solutions for values of $\lambda\leq\omega^2$ which 
are well 
behaved at $r=0$ and $r=\infty$.
Therefore the continuous spectrum of eigenvalues can be written as 
$\lambda=\omega^2+
\omega^2_\lambda$
and:
\beq
\det[-{\cal D}_n+\omega^2]=\prod (\omega^2+\omega^2_\lambda).
\eeq

The numerator of (\ref{ratio}) involves the ``potential" $U''(\phi_b(r))$.
\begin{figure}
\epsfysize=3.5in
\epsfbox[-10 424 481 716]{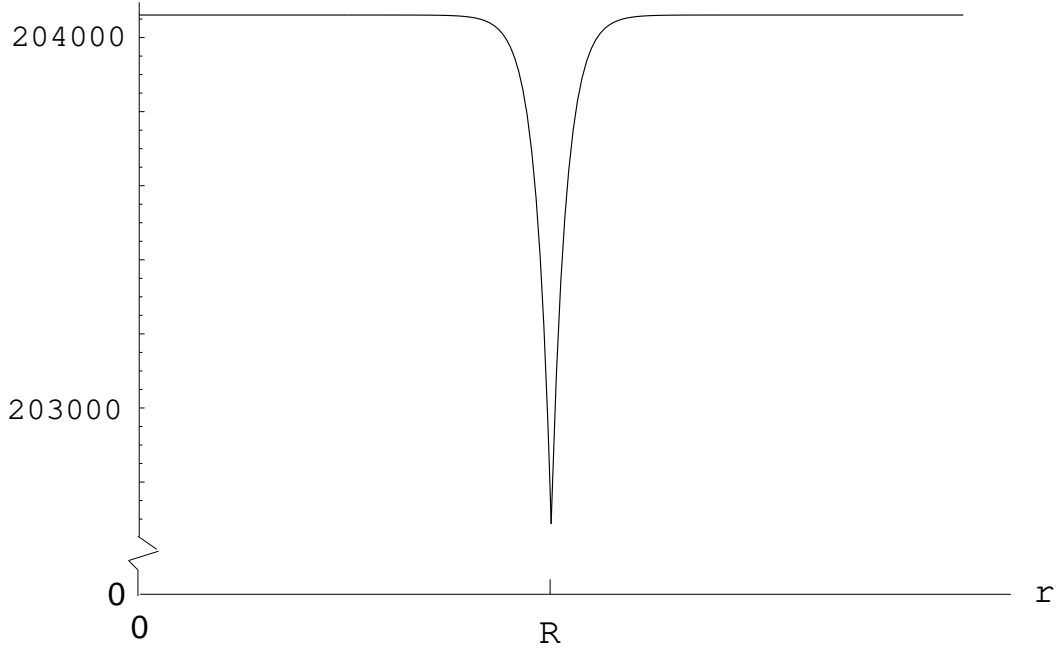}
\caption{The ``potential'' $U''(\phi_b(r))$.} 
\label{quantumpotential}
\end{figure}
For the symmetrized effective potential (\ref{symmetricpotential}) the
``potential" for this problem is approximately constant except for a
small perturbation in a small region near the bubble wall (see
Fig.\ref{quantumpotential}).  The constant potential has been analyzed
above.  The solutions with or without the perturbations are identical
for $r\ll R$ and almost identical for $r\gg R$.  The perturbed
solution differs from the unperturbed solution in this latter region
at most by normalization and a phase shift, $\tilde{\omega}_\lambda r
= \omega_\lambda r + \delta(\omega_\lambda)$.  In this situation, for
each value of n, we can obtain the ratio of determinants for a
discrete spectrum by \cite{VZNS}:
\beq
\prod \frac{\omega^2 + \tilde{\omega}_\lambda^2}{\omega^2 + \omega_\lambda^2}
=\exp\left(
 \sum \ln \frac{\omega^2 + \tilde{\omega}_\lambda^2}{\omega^2 + 
\omega_\lambda^2} \right)
 \approx 
\exp\left( \sum \frac{2 \omega_\lambda (\tilde{\omega}_\lambda-\omega_\lambda)}
{\omega^2
 + \omega_\lambda^2} \right),
\eeq
which becomes in the continuum:
\beq
\exp \left[ \frac{1}{\pi}\int_0^\infty d\omega_\lambda  \frac{2\omega_\lambda 
\delta(\omega_\lambda)}{\omega^2 + \omega_\lambda^2} \right].
\label{detratio}
\eeq
This formula gives the determinant ratio for a particular value of `n'
if we know the phase shifts, $\delta(\omega_\lambda)$.  In order to
calculate the phase shifts, however, we must make further
approximations.

Consider both the perturbed and the unperturbed equations:
\bea
\left[ \frac{d^2}{dr^2}- \frac{4n^2+8n+3}{4r^2} + (\lambda-\omega^2) \right] 
u(r)=0 
\label{unpert},\\
\left[ \frac{d^2}{dr^2}- \frac{4n^2+8n+3}{4r^2} + \left(\lambda-\omega^2+
V_{\mbox{pert}}
(r)\right)\right] v(r)=0.
\label{pert}
\eea
Multiply (\ref{unpert}) by $v(r)$ and (\ref{pert}) by $u(r)$,
subtract, and integrate from 0 to $\infty$ to obtain:
\beq
\int_0^\infty \left( u''(r) v(r) - v''(r) u(r) \right) dr= \left( u'(r)v(r)-
v'(r)u(r) 
\right)|_0^\infty = \int_0^\infty V_{\mbox{pert}}(r) u(r) v(r) dr.
\eeq
Using the fact that both solutions vanish at $r=0$ and using the asymptotic
form of the Bessel functions we obtain an exact formula for the phase
shift, $\delta(\omega_\lambda)$:
\beq
\sin \delta(\omega_\lambda)= \frac{1}{\omega_\lambda} \int_0^\infty V_{\mbox{pert}}(r) u(r) 
v(r) dr.
\label{sindelta}
\eeq
Of course, since the perturbed differential equation is extremely
difficult to solve, we use perturbation theory to obtain:
\beq
\delta(\omega_\lambda)\approx\delta(\omega_\lambda)_0= \frac{1}{\omega_\lambda} \int_0^\infty V_{\mbox{pert}}
(r) 
u(r)^2 
dr = 
\frac{\pi}{2}\int_0^\infty V_{\mbox{pert}}(r) r J_{n+1}(\omega_\lambda r)^2 dr,
\label{delta}
\eeq
for small phase shifts. Using this result for the phase shift in
Equation (\ref{detratio}), we obtain the ratio of determinants:
\beq
\left|\frac{-\det[{\cal D}_n+U''(\phi_b)]}{\det[-{\cal D}_n+\omega^2]} 
\right|
\approx \exp \left[ \int_0^\infty\int_0^\infty d\omega_\lambda dr \frac{ 
\omega_\lambda 
r}{\omega^2 + \omega_\lambda^2}   V_{\mbox{pert}}(r) J_{n+1}
(\omega_\lambda r)^2 
 \right].
\label{detratio2}
\eeq
for each value of n.
\begin{figure}
\epsfysize=3.5in
\epsfbox[-10 377 538 691]{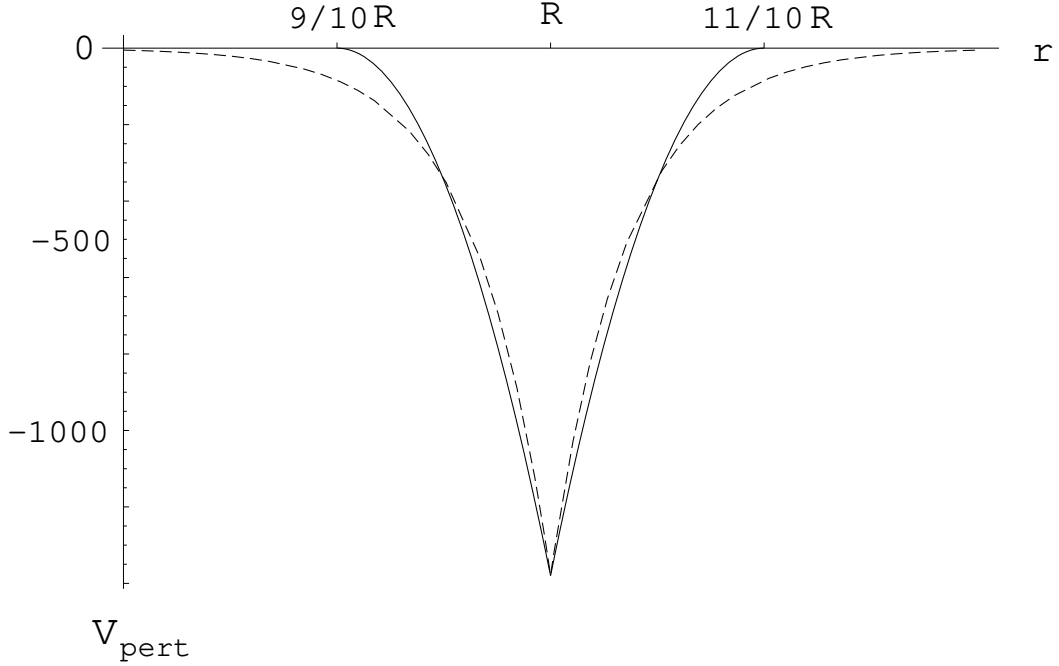}
\caption{The perturbation $V_{pert}(r)$ is the dotted line and the
 approximate perturbation $\tilde{V}_{pert}(r)$ is the solid line.} 
\label{perturbation}
\end{figure}
The complete determinant ratio then becomes:
\beq
 \left|\frac{\det[-\partial_\mu
 \partial^\mu+U''(\phi_b)]}{\det[-\partial_\mu
 \partial^\mu+\omega^2]} \right| = \exp
 \sum_{n=0}^\infty(2n(n+2)+1)\left[ \int_0^\infty\int_0^\infty
 d\omega_\lambda dr
\frac{ \omega_\lambda r}{\omega^2 + \omega_\lambda^2}   V_{\mbox{pert}}(r) 
J_{n+1}(
\omega_\lambda r)^2  \right].
\label{completedetratio}
\eeq
for an arbitrary perturbation potential.  The next step in our
analysis is to take into account the specific potential in our
problem.

It should be noted that this approximation gives the same result
as the exact answer expanded to one loop (see Eq.\ref{loopexpansion}
and \ref{1loop}) where the trace is taken instead over 
$|r n l m \rangle$, $|k n l m \rangle$ bases.

This formula is suitable for numerical evaluation, but in order to obtain
an analytical answer we must approximate the perturbation in the potential
$U''$ by:
\beq
\tilde{V}_{\mbox{pert}}(r)= \left\{
  \begin{array}{l}
  -\beta \frac{10 (r-\frac{11R}{10})^2}{R^2}~~R>r>
     \frac{11R}{10} \\
    -\beta\frac{10 (r-\frac{9R}{10})^2}{R^2}~~\frac{9R}{10}>r>R \\
 ~~~~~~~ 0~~~~~~~~~~~\mbox{otherwise}
  \end{array},
\right.
\eeq
where $\beta=\left(U''(\phi(0))-U''(\phi(R))\right)$(see Fig. 
\ref{perturbation}).
With this approximate potential we find the phase shift via
(\ref{delta}):
\bea
\delta_0(\omega_\lambda)=-200\frac{ \beta}{R^2}\left[ {\frac{121}{200}r^2 R^2 \left(J_{n+1}
(r\omega_\lambda )^2 -
  J_n(r\omega_\lambda)
          J_{n+2}(r\omega_\lambda) \right) } \right. ~~~~~~~~~~~~~~~~~~~~~~~
~~~~
~~~~~~~~~~  
\nonumber\\
+\frac{2^{-4-2n}r^4(r\omega_\lambda)^{2+2n}}{(n+4)!(n+1)!} \left\{
2(n+2)(n+4)\;{}_1F_2\left(n+\frac{3}{2},n+4,2n+3;-r^2\omega_\lambda^2\right)
\right. \nonumber \\
\left. -r^2\omega_\lambda^2 \;{}_1F_2\left(n+\frac{5}{2},n+5,2n+4;-r^2
\omega_\lambda^2\right)\right\} 
\nonumber \\
\left. \frac{11}{10}\frac{2^{-2-2n}r^3 R(r\omega_\lambda)^{2+2n}}{(n+\frac{5}
{2})
[(n+1)!]^2} \;
{}_2F_3\left(n+\frac{3}{2},n+\frac{5}{2},n+2,n+\frac{7}{2},2n+3;-r^2
\omega_\lambda^2\right)
\right]^{11/10 R}_R.
\label{exactdelta}
\eea

\begin{figure}
\epsfysize=3.5in
\epsfbox[83 439 551 701]{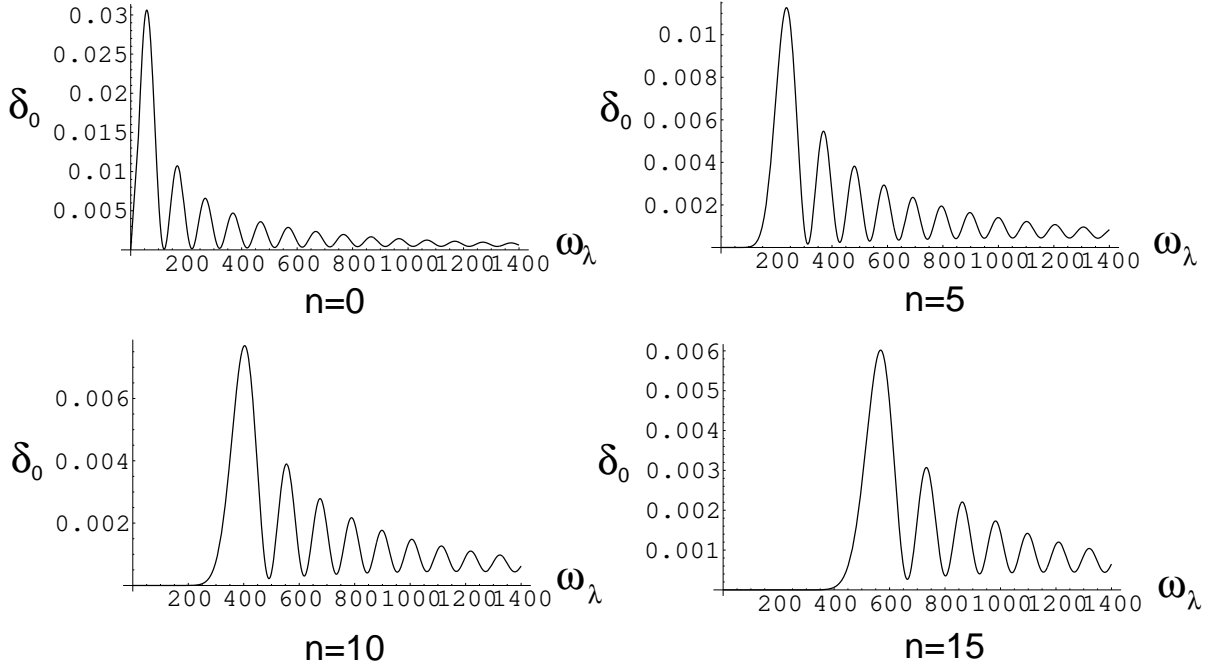}
\caption{The phase shift $\delta_0(\omega_\lambda)$ shown as a function of 
$\omega_\lambda$ for various 
values of n.  Notice the changing vertical scale.} 
\label{phaseshift}
\end{figure}
Using this result we can verify that $\delta(\omega_\lambda) \ll 1$ which means that
the approximations of (\ref{delta}) are justified.  The phase shift as
a function of $\omega_\lambda$ is shown for a few values of n in
Fig.(\ref{phaseshift}).  However, (\ref{exactdelta}) is very difficult
to work with so we make another approximation.  For $\omega_\lambda$
large enough we can use the asymptotic approximation for the Bessel
function in (\ref{delta}),
\beq
J_{n+1}(x)=\sqrt{\frac{2}{\pi x}} \left\{P_{n+1}(x)\cos\left[
x-\left(n+\frac{3}{2}\right) \frac{\pi}{2}\right]-Q_{n+1}(x)\sin\left[
x-\left(n+\frac{3}{2}\right) \frac{\pi}{2}\right] \right\},
\label{Besselapprox}
\eeq
where:
\bea
P_{n+1}(x)&=& 1 - \frac{(\mu-1)(\mu-9)}{2!\; (8x)^2}+\frac{(\mu-1)(\mu-9)
(\mu-25)
(\mu-49)}{4!\; (8x)^4}- \cdots, \\
Q_{n+1}(x)&=& \frac{(\mu-1)}{1!\; (8x)}-\frac{(\mu-1)(\mu-9)(\mu-25)}{3!\; 
(8x)^3} + 
\cdots, \\
\mu&=& 4(n+1)^2.
\eea
In this way we find:
\bea
\delta_0(\omega_\lambda)&\approx& \frac{2R\beta}{30 \pi \omega_\lambda} + \frac{(-1)^n\beta}
{\pi} 
\frac{\cos(2R\omega_\lambda )}{\omega_\lambda^2}
 + {\cal O}(1/\omega_\lambda^3).
\label{deltaapprox}
\eea

Of course this estimate for $\delta_0(\omega_\lambda)$ has an infrared cutoff in
$\omega_\lambda$ below which it is not a good estimate and this will
be true no matter how many terms in the approximation of the Bessel
function we keep.  Therefore as a first approximation we keep only the
first term and cut off the $\omega_\lambda$ integral at the position
of the first peak in $\delta_0(\omega_\lambda)$. Using this
approximation for $\delta_0(\omega_\lambda)$ we obtain our first
estimate of the ratio of determinants for each value of the quantum
number 'n' given by:
\beq
\frac{\det[-{\cal D}_n+U''(\phi_b)]}{\det[-{\cal D}_n+U''(\phi_-)]} 
\approx 
\exp\left[
\frac{2R\beta}{30 \pi \omega} \arctan \left(\frac{\omega}{\bar{\omega}(n)}
\right)
 \right],
\label{deltan}
\eeq
where $\bar{\omega}(n)\approx 33.16n+72=a n + b$ is the infrared
cutoff given by the location of the first peak in
$\delta_0(\omega_\lambda)$.  Using the result (\ref{deltan}) we can
calculate our first approximation for the complete determinant ratio.
For large values of n the terms of the sum in (\ref{completedetratio})
approach:
\beq
\frac{4R\beta}{30 \pi \omega} n^2 \left(\frac{\omega}{an}\right),
\eeq
and therefore the sum is infinite.  Our first estimate for the
complete determinant ratio based on (\ref{completedetratio}) is
infinite.  This was to be expected based on comments at the beginning
of this section and, since this approximation exactly coincides with
the tadpole diagram, we subtract this term from the exponent as our
normalization prescription for the complete determinant ratio.  Note
that we have used the approximation of (\ref{deltaapprox}) to
calculate the infinite factor but the equality with the tadpole
diagram holds before this approximation was made.
 
In order to find a finite result we must adjust our earlier
approximation for the phase shift:
\beq
\delta(\omega_\lambda)=\arcsin(\delta_0(\omega_\lambda))= \delta_0(\omega_\lambda) + \frac{1}{6} \delta_0(\omega_\lambda)^3 + \frac{3}{40} 
\delta_0(\omega_\lambda)^5+
 \cdots
\label{sindeltaapprox}
\eeq
which would add correction terms to the exponent in (\ref{deltan}).
The second term in (\ref{sindeltaapprox}) leads to a factor of the form:
\beq
\exp \left[ \frac{\pi^2}{24}\int_0^\infty d\omega_\lambda  
\frac{\omega_\lambda \delta_0(\omega_\lambda)^3}
{\omega^2 + \omega_\lambda^2} \right],
\label{newdetratio}
\eeq
for each value of n. This term is also divergent but does \textbf{not} exactly
coincide with the next term in the expansion (\ref{loopexpansion}). 
However, the divergent contribution in each must be the same.

Therefore dividing through by the previously obtained infinite factor ({\it ie.} the right hand side of (\ref{deltan})) leads to the renormalized value: 
\beq
\frac{\det[-{\cal D}_n+U''(\phi_b)]}{\det[-{\cal D}_n+U''(\phi_-)]} 
\approx
\exp\left[ \frac{\pi^2}{24}
  \left(\frac{2R\beta}{30 \pi \omega} \right)^3 \left\{
\frac{\omega}{\bar{\omega}(n)}-
\arctan \left(\frac{\omega}{\bar{\omega}(n)}\right) 
\right\}\right],
\label{nextratio}
\eeq
For large values of $n$ the terms of the sum in
(\ref{completedetratio}) approach:
\beq
\frac{\pi^2}{24}\left(\frac{2R\beta}{30 \pi \omega}\right)^3 (2n^2+4n+1) 
\left(\frac{\omega^3}{3a^3n^3}\right).
\eeq
which also leads to a divergent sum and therefore an infinite value
for the determinant ratio. The divergence here must be contained in
the divergence in the two interaction term of the expansion
(\ref{loopexpansion}), as are other logarithmic divergences obtained
from keeping more terms in the approximation of the Bessel function
(\ref{Besselapprox}).  However, subtracting the two interaction term
will almost certainly leave a finite contribution at the next order in
$n$.  We will assume this is the case but since we are doing an
approximate calculation we will not calculate the finite contribution
from the two interaction term or any correction terms to our
approximation since they will have the same physical dependence as the
approximation we will give.  We should further note that correction
terms are probably not calculable analytically and are not likely to
be a problem for our results.  They are obtained from integrating
oscillatory functions (see second term in (\ref{deltaapprox})) over
$\omega_\lambda$ in (\ref{newdetratio}) and should not contribute very
much.  The renormalized approximation to the complete determinant
ratio (\ref{completedetratio}) that we obtain from (\ref{nextratio})
is given by:
\bea
\exp  \frac{\pi^2}{24}\left(\frac{2R\beta}{30 \pi \omega} \right)^3 
\left[\sum_{n=0}^\infty(2n(n+2)+1) 
  \left\{
\frac{\omega}{\bar{\omega}(n)}-\arctan \left(\frac{\omega}{\bar{\omega}(n)}\right) \right\}   - \sum_{n=1}^\infty 2n^2\frac{\omega^3}{3a^3n^3} \right]
\eea
\bea
 =\exp  \frac{\pi^2}{24}\left(\frac{2R\beta}{30 \pi \omega} \right)^3 
\left[\sum_{n=0}^{100}(2n(n+2)+1) 
  \left\{
\frac{\omega}{\bar{\omega}(n)}-\arctan \left(\frac{\omega}{\bar{\omega}(n)}\right) \right\}\right. \nonumber \\
\left. -\sum_{n=1}^{100}\left\{(2n(n+2)+1)\frac{\omega^3}{3a^3n^3}\right\}+\frac{4\omega^3}{3a^3} \zeta(2) +\frac{\omega^3}{3a^3} \zeta(3)
\right].
\eea
Evaluating gives:
\beq
\exp \left[-5300 \frac{\pi^2}{24}\left(\frac{2R\beta}{30 \pi \omega} 
\right)^3 
\right]\approx \exp[-2\times10^{-5}].
\eeq
Because of the small value of the exponent we can see that the exponential
is extremely well approximated by:
\beq
\exp \left[ \frac{-5300\pi^2}{24}\left(\frac{2R\beta}{30 \pi \omega} 
\right)^3 
\right]\approx 1 - \frac{5300\pi^2}{24}\left(\frac{2R\beta}{30 \pi 
\omega} \right)^3.
\eeq
which is our result for the determinant ratio (\ref{ratio}).

This is a nonperturbative calculation because Eq.(\ref{detratio}) is
a nonperturbative resummation of the perturbation expansion
(\ref{loopexpansion}) of the determinant.  While we make an approximation through
Eq.(\ref{sindeltaapprox}) this expansion is nonperturbative since the
terms of the expansions do not coincide.  Our approximate calculation
should therefore have contributions from all perturbative diagrams.
This nonperturbative resummation of diagrams in the determinantal prefactor is
very similar in spirit to that of \cite{Kiers}.

This result constitutes the contribution of the positive eigenvalues
to the quantum corrections to the decay rate in the zero temperature
theory.  In the next section we consider the zero and negative
eigenvalues.

\subsection{Zero and Negative Eigenvalues}

The zero eigenvalues contribute $\sqrt{B/2\pi}$ per collective
coordinate\cite{Okun}.  The action of the 4D bubble is independent of
the center of the bubble which means there are 4 collective
coordinates leading to the first factor in Eq.(\ref{A}).

The eigenfunctions of zero eigenvalue are:
\beq
\chi_{o\mu}(x,y,z,t)= \frac{d}{dx^\mu} \phi_b(r)= \frac{dr}{dx^\mu} 
\;\frac{d}{dr}\phi_b(r) =\frac{x^\mu}{r} \;
 \frac{d}{dr}\phi_b(r),
\eeq
where $r=\sqrt{x^2+y^2+z^2+t^2}$ and:
\bea
t=& r \; Cos(\psi)& \sim Y_{100}, \\
z=& r \;Cos(\theta) Sin(\psi)& \sim Y_{110}, \\
y=& r \;Cos(\phi) Sin(\theta) Sin(\psi) & \sim Y_{111}+ Y_{11-1},\\
y=& r \; Sin(\phi) Sin(\theta) Sin(\psi) & \sim Y_{111}- Y_{11-1}.
\eea
Therefore these eigenfunctions all correspond to $n=1$ and since there
are no radial nodes we can be sure that there are no negative eigenvalues
with $n\neq 0$.

In \cite{Aspects}, Coleman argued that there is only a single negative
eigenvalue for an O(4) invariant bounce.  We assume this\footnote{This
is most likely a good assumption but not proven due to the cusp in our 
potential.  For more
details see \cite{Callan},\cite{Aspects} and \cite{CGM}} and determine its value.
We use the method of
\cite{Okun} with a slight modification.  As was argued by Coleman, the
only possible eigenfunctions of negative eigenvalue are those that are
bound to the bubble wall.  For such eigenfunctions we can approximate
the centrifugal potential in (\ref{laplacian}) by a constant
determined by its value at the bubble wall ($r=R$):
\beq
\lambda_{pn}=\lambda_p + \frac{4n^2+8n+3}{4R^2},
\eeq
where $\lambda_p$ is a number independent of $n$.
We know that for $n=1$ the lowest eigenvalue is zero:
\beq
\lambda_{01}=\lambda_0 + \frac{15}{4R^2}=0.
\eeq
Therefore we can obtain the lowest eigenvalue for $n=0$:
\beq
\lambda_-=\lambda_{00}=\lambda_0 + \frac{3}{4R^2}=-\frac{15}{4R^2}+ \frac{3}{4R^2}=
- \frac{3}{R^2}.
\eeq
This value is different from Coleman's but only by a factor of 2.
We cannot explain this discrepancy but can only stand by our calculation.

\subsection{Decay Rate for Zero Temperature}
Therefore combining the zero temperature semiclassical result
(\ref{19f}) with the quantum corrections obtained in the last section we
estimate the decay rate per unit volume of the false vacuum in the
zero temperature theory to be:
\bea
\Gamma/V&=& \exp\left(-S_4 \right) \left(\omega^4\left(\frac{S_4}{2 \pi}\right)^2\right)
  \frac{\omega} {\sqrt{\lambda_-}}
 \left|\frac{\det'[-\partial_\mu \partial^\mu+U''(\phi_b)]}{(\omega^{-2})^5\det[-\partial_\mu
 \partial^\mu+\omega^2]} \right| \\
&=& \exp \left( -\frac{27\pi^2\sigma^4}{2(\Delta E)^3} \right)
\left( \frac{27\pi^2\sigma^4\omega^2}{4 \pi(\Delta E)^3}\right)^2  \frac{R\omega}{\sqrt{3}}
\left[ 1 - \frac{5300\pi^2}{24}\left(\frac{2R\beta}{30 \pi \omega} 
\right)^3 \right]^{-1/2} \\
&=& 1.55\times 10^{-18} \mbox{MeV}^4\approx3\times 10^{-4} \mbox{fm}^{-3}
\mbox{s}^{-1}.
\eea
It should be noted that the quantum corrections are negligible so the
decay rate is basically determined by the semiclassical result.  We
believe that this observation would be unchanged by the inclusion of
corrections that we have ignored in this calculation.  While the
quantum corrections did not turn out be significant in this case, the
fact that they are not is relevant to the baryogenesis mechanism
described in \cite{BHZ}.  As well, the techniques
applied to the problem may prove useful in other calculations of this
type.

In the next section we perform the same calculation in the high temperature
limit.

\section{Quantum Corrections for High Temperature}
The quantum corrections to the decay rate at high temperature
correspond to\cite{Linde2}:
\beq
A=T\left(\omega^3 \int d^3x\left( \frac{S_3[\phi_b]}{2\pi T}\right)^{3/2}\right)
\frac{\omega} {\sqrt{\lambda_-}}
 \left|\frac{\det'[
-\partial_i \partial^i
+U''(\phi_b)]}{(\omega^{-2})^3\det[-\partial_i \partial^i+U''(\phi_-)]} \right|^{-1/2} .
\label{Ahigh}
\eeq
which again factors into three parts corresponding to zero, negative
and positive eigenvalues respectively.

\subsection{Positive Eigenvalues}
The calculation in three dimensions is extremely similar to the four dimensional
calculation presented in the Sect. 4.  The
expansion of the determinant ratio is exactly the same as in 4D
(\ref{loopexpansion}).  Tracing over Cartesian bases we can see that
the one loop term is divergent while the two loop term is finite:
\bea
Tr \left[ \frac{V_{pert}(r)}{-\partial_\mu \partial^\mu+\omega^2} \right]&=&
\int
 d^3x \;d^3k \; \frac{V_{pert}(r)}{k^2+\omega^2} \label{1loophigh},\\ 
Tr \left[ \left(\frac{V_{pert}(r)}{-\partial_\mu \partial^\mu+\omega^2} 
\right)^2
\right]&=& \int \;d^3k \;d^3p
 \; \frac{{\cal V}_{pert}(k)}{(k^2+\omega^2)} \frac{{\cal V}_{pert}(-k)}{(
(k+p)^2+
\omega^2)}.
\label{2loophigh}
\eea

The calculation is done using spherical coordinates and the eigenfunctions
are expanded in terms of the spherical harmonics:
\beq
\chi(r,\theta,\phi)=
\sum_{n=0}^\infty \sum_{l=-n}^{n} \sum_{m=-|l|}^{|l|} C_{lm}\frac{u(r)}{r} 
Y_{lm}(\theta,\phi).
\label{expansion3D}
\eeq
For each value of $l$ the Laplacian becomes:
\beq
\partial_i \partial^i  \rightarrow \frac{1}{r} \left[ \frac{d^2}{dr^2}- 
\frac{l(l+1)}{r^2} \right]r \equiv \frac{1}{r} {\cal D}_l r.
\label{laplacianhigh}
\eeq
Therefore:
\beq
\det[-\partial_i \partial^i+U'']=\prod_{l=0}^\infty 
(\det[-{\cal D}_l+U''])^{2l+1}.
\label{lprod}
\eeq
The solutions to this differential equation that are well behaved at 
$r=0$ and $r=\infty$ are:
\beq
u(r)= \omega_\lambda r j_l(\omega_\lambda r)\equiv \sqrt{\frac{\pi
\omega_\lambda 
r}{2}}
 J_{l+1/2}(\omega_\lambda r),
\eeq
where $j_l$ are the usual spherical Bessel functions and $\omega_\lambda\equiv 
\sqrt{\lambda-\omega^2}$.
We obtain:
\beq
\delta(\omega_\lambda)\approx \frac{1}{\omega_\lambda} \int_0^\infty V_{\mbox{pert}}(r) u(r)^2
 dr
 = 
\frac{\pi}{2}\int_0^\infty V_{\mbox{pert}}(r) r J_{l+1/2}(\omega_\lambda r)^2 
dr,
\eeq
thus giving\footnote{Note that we have again dropped the factor of 
$(\omega^{-2})^3$ and the ``det$^\prime$'' notation as the omitted eigenvalues 
correspond to a set of measure zero.}
\footnote{The summand in this expression appears in \cite{Cottingham} but is only 
used for the high energy modes.  The non-divergent term in what follows does not appear in 
\cite{Cottingham}.}:
\beq
 \left|\frac{\det[-\partial_i
 \partial^i+U''(\phi_b)]}{\det[-\partial_i
\partial^i+\omega^2]}
 \right| = \exp \sum_{l=0}^\infty(2l+1)\left[
 \int_0^\infty\int_0^\infty d\omega_\lambda dr \frac{ \omega_\lambda
 r}{\omega^2 + \omega_\lambda^2} V_{\mbox{pert}}(r)
 J_{l+1/2}(\omega_\lambda r)^2 \right].
\label{completedetratiohigh}
\eeq
As in the zero temperature case we can find the exact expression for
the phase shift with the approximate potential but it is too difficult
to work with.  Instead we approximate the Bessel functions as in
(\ref{Besselapprox}) where $(n+1) \rightarrow (l+1/2)$.
In this way we find:
\bea
\delta(\omega_\lambda)&\approx&\delta_0(\omega_\lambda)= \frac{2R\beta}{30 \pi \omega_\lambda}
 + {\cal O}(1/\omega_\lambda^2).
\eea
Our first estimate of the ratio of determinants for each value of the
quantum number 'l' is given by:
\beq
\frac{\det[-{\cal D}_l+U''(\phi_b)]}{\det[-{\cal D}_l+U''(\phi_-)]} 
\approx 
\exp\left[
\frac{2R\beta}{30 \pi \omega} \arctan \left(\frac{\omega}{\bar{\omega}(n)}
\right)
 \right],
\eeq
where $\bar{\omega}(l)\approx 49.5l+82=c l + d$ is the infrared cutoff
given by the location of the first peak in $\delta_0(\omega_\lambda)$
as a function of $\omega_\lambda$.  For large values of $l$ the terms
of the sum in (\ref{completedetratiohigh}) approach:
\beq
\frac{4R\beta}{30 \pi \omega} l \left(\frac{\omega}{cl}\right),
\eeq
and the sum is infinite and therefore the complete determinant ratio
is also infinite.  Again this was to be expected and, since this
approximation exactly coincides with the one loop term, our
renormalization prescription is to remove the one loop term from the exponent of
 (\ref{completedetratiohigh}).
 
Again we must adjust our earlier approximation for the phase shift:
\beq
\delta(\omega_\lambda)=\arcsin(\delta_0(\omega_\lambda))= \delta_0(\omega_\lambda) + \frac{1}{6} \delta_0(\omega_\lambda)^3 + \frac{3}{40} 
\delta_0(\omega_\lambda)^5
+ \cdots 
\eeq
and the correction is implemented as in (\ref{newdetratio}).  Evaluating
the contribution to the determinant ratio leads to:
\beq
\frac{\det[-{\cal D}_l+U''(\phi_b)]}{\det[-{\cal D}_l+U''(\phi_-)]} 
\approx 
\exp\left[ \frac{\pi^2}{24}
  \left(\frac{2R\beta}{30 \pi \omega} \right)^3 \left\{
\frac{\omega}{\bar{\omega}(l)}-\arctan 
\left(\frac{\omega}{\bar{\omega}(l)}\right) 
\right\}\right].
\eeq
For large values of $l$ the terms of the sum in
(\ref{completedetratiohigh}) approach:
\beq
\frac{\pi^2}{24}\left(\frac{2R\beta}{30 \pi \omega}\right)^3 (2l+1) 
\left(\frac{\omega^3}{3c^3l^3}\right).
\eeq
The sum is finite so we do not need to further normalize and our
result for the complete determinant ratio (\ref{completedetratiohigh}) is:
\bea
 \exp \frac{\pi^2}{24}\left(\frac{2R\beta}{30 \pi \omega} \right)^3
 \left[\sum_{l=0}^{100}(2l+1) \left\{
\frac{\omega}{\bar{\omega}(l)}-
\arctan
 \left(\frac{\omega}{\bar{\omega}(l)}\right)
 \right\}\right. \nonumber \\
\left. -\sum_{l=1}^{100}\left\{(2l+1)\frac{\omega^3}{3c^3l^3}\right\}+\frac{2\omega^3}{3a^3} 
\zeta(2) 
+\frac{\omega^3}{3a^3} \zeta(3) \right].
\eea
Evaluating gives:
\beq
\exp \left[82 \frac{\pi^2}{24}\left(\frac{2R\beta}{30 \pi \omega} 
\right)^3 
\right]\approx \exp[3\times10^{-7}].
\eeq
Because of the small value of the exponent we can see that the exponential
is extremely well approximated by:
\beq
\exp \left[ \frac{82\pi^2}{24}\left(\frac{2R\beta}{30 \pi \omega} 
\right)^3 
\right]\approx 1 + \frac{82\pi^2}{24}\left(\frac{2R\beta}{30 \pi 
\omega} \right)^3.
\eeq
This result constitutes the contribution of the positive eigenvalues
to the quantum corrections to the decay rate in the high temperature
theory.  In the next section we consider the zero and negative
eigenvalues.

\subsection{Zero and Negative Eigenvalues}
In the O(3) invariant bubble there are 3 collective coordinates leading to the
first term in (\ref{Ahigh}).
The eigenfunctions of zero eigenvalue are:
\beq
\chi_{o\mu}(x,y,z)= \frac{d}{dx^\mu} \phi_b(r)= \frac{dr}{dx^\mu} 
\;\frac{d}{dr}\phi_b(r) =\frac{x^\mu}{r} \;
 \frac{d}{dr}\phi_b(r),
\eeq
where $r=\sqrt{x^2+y^2+z^2}$ and:
\bea
z=& r \; Cos(\theta)& \sim Y_{10}, \\
y=& r \; Cos(\phi) Sin(\theta) & \sim Y_{11}+ Y_{1-1},\\
y=& r \; Sin(\phi) Sin(\theta) & \sim Y_{11}- Y_{1-1}.
\eea
Therefore these eigenfunctions all correspond to $l=1$ and since there
are no radial nodes we can be sure that there are no negative
eigenvalues with $l\neq 0$.

We will assume, as we did in the previous section, that there is only a
single negative eigenvalue and concentrate on obtaining an
approximation.  The only possible eigenfunctions of negative
eigenvalue are those that are bound to the bubble wall.  For such
eigenfunctions we can approximate the centrifugal potential in
(\ref{laplacianhigh}) by a constant:
\beq
\lambda_{pn}=\lambda_p + \frac{l(l+1)}{R^2},
\eeq
where $\lambda_p$ is a number independent of $l$.
We know that for $l=1$ the lowest eigenvalue is zero:
\beq
\lambda_{01}=\lambda_0 + \frac{2}{R^2}=0.
\eeq
Therefore we can obtain the lowest eigenvalue for $l=0$:
\beq
\lambda_-=\lambda_{00}=\lambda_0 =-\frac{2}{R^2}.
\eeq

\subsection{Decay Rate for High Temperature}
Therefore combining the high temperature semiclassical result
(\ref{HighTSemiclassical}) with the quantum corrections obtained in
the last section we estimate the decay rate per unit volume of the
false vacuum in the high temperature theory to be:
\beq
\Gamma/V= \frac{R\omega}{\sqrt{2}} \left( \omega^3\left( \frac{16\pi S_1(T)^3}
{3(\Delta E)^2 T}
\right)^{3/2} \right) \exp \left( -\frac{16\pi S_1(T)^3}
{3(\Delta E)^2 T}
\right)
\left[ 1 + \frac{82\pi^2}{24}\left(\frac{2R\beta}{30 \pi \omega} 
\right)^3 \right]^{-1/2}.
\label{decayratehighT}
\eeq
We cannot obtain a numerical estimate for the decay rate at this time
since the temperature dependence of the semiclassical result is not
currently known.  This result could be important to the study of
false vacuum states of this type in heavy ion collisions.

\section{Conclusion}
We have obtained an estimate for the decay rate per unit volume of
$\Gamma/V\approx3\times 10^{-4} \mbox{fm}^{-3} \mbox{s}^{-1}$ for a
false vacuum in an effective Lagrangian approach to QCD for zero
temperature theory.  We have obtained an expression
(\ref{decayratehighT}) for the decay rate per unit volume in the high
temperature theory, which is the best we can do without knowledge of
the temperature dependence of the effective potential.  These are
nonperturbative calculations in that they contain contributions from 
all orders of perturbation theory.

The value of the decay rate for zero temperature shows that if the
universe had started out in a false vacuum state then it would have
decayed long ago into the true vacuum state.  This is not new or
interesting.  The interesting thing concerns the possibility
that these vacuum bubbles would have significant effects on the
evolution of the early universe at around the time of the QCD phase transition.

One such effect related to baryogenesis was described in \cite{BHZ}.
The first approximation to the decay rate we have calculated shows that
bubbles of false vacuum are far too unstable for this simplified baryogenesis
mechanism.  The effect of fermions
or intrinsic heavy degrees of freedom could go a long way to
stabilizing the bubbles because they would make the barrier separating
the false vacuum from the true vacuum much higher, thus increasing the
stability of the false vacuum.  The precise calculation is outside the
scope of the present work and we can only say that, so far, this
mechanism for baryogenesis at the QCD scale has not been proved
viable.  

The result does not, however, rule out the possibility of nontrivial
effects on the evolution of the early universe.  As well, the techniques
developed may be useful in the study of arbitrary metastable vacua which
seem to be a general feature of gauge theories in the strong coupling regime.

The high temperature expression for the decay rate will be relevant to
the possibility of observing CP odd bubbles at RHIC through signatures
as described in \cite{Kharzeev}.  As was mentioned previously the
false vacuum described in \cite{Kharzeev} assumed the large $N_c$
limit.  However, the same effects would manifest themselves in our
case which arises from an effective potential valid for arbitrary
$N_c$

In the future it might be useful to verify numerically our contention
that the corrections neglected in the approximation and
finite contributions from the renormalization prescription do not
significantly alter the results.  As well, one could do numerical
calculations using the exact numerical domain wall solution in the
case of non-degenerate vacuum states.  It is possible that the
determinant could also be estimated numerically in the non-degenerate
case.

It would also be interesting to repeat this calculation using the methods of
\cite{Cottingham} \cite{Kiselev} \cite{Baacke} \cite{Strumia} and compare
with the present results.

\appendix

\section{Hyperspherical Harmonics in Four Dimensions}

Hyperspherical coordinates in 4D dimensions are related to the
Cartesian coordinates by: 
\bea &&x=r \; Sin[\psi] Sin[\theta] Sin[\phi],\\ 
&&y=r \;Sin[\psi] Sin[\theta] Cos[\phi] ,\\
&&z=r \;Sin[\psi] Cos[\theta], \\
&&w=r \;Cos[\psi].
\eea 
The Laplacian in 4D in
hyperspherical coordinates is: 
\beq 
\frac{1}{r^3} \partial_r \left(r^3
\partial_r \right) + \frac{1}{r^2 \sin^2 \psi} \partial_\psi
\left(\sin^2 \psi \partial_\psi \right)+ \frac{1}{r^2 \sin^2 \psi
\sin\theta} \partial_\theta \left(\sin\theta \partial_\theta \right)+
\frac{1}{r^2 \sin^2\psi \sin^2 \theta } \partial^2_\phi. 
\eeq 
Assuming separable solutions and treating $\theta$ and $\phi$ 
coordinates in exactly the same way as in three dimensions we obtain
the differential equation:
\beq
\Psi^{\prime\prime}(\psi)+ 2 \cot(\psi)\Psi^{\prime}(\psi)-l(l+1) \csc^2(\psi) 
\Psi(\psi) = \Lambda \Psi(\psi).
\eeq
With the substitution $u=Cos[\psi]$ this becomes:
\beq
(1-u^2) U^{\prime\prime}(u) -3 u U^{\prime}(u)- \frac{l(l+1)}{1-u^2}U(u)= B 
U(u).
\eeq
If $B=n(n+2)$ and $l(l+1)=l^{\prime}(l^{\prime}+1)$ this can be
identified as the differential equation satisfied by the associated
type II Chebyshev functions.  These can be obtained from the type II
Chebyshev differential equation in exactly the same way as associated
Legendre functions are obtained from the Legendre differential
equation.
  
As an aside notice that the type II Chebyshev equation is a special
case of the Geigenbauer (Ultraspherical) equation:
\beq
(1-x^2) \frac{d^2}{dx^2}C_n^{(\alpha)}(x) -(2 \alpha +1) x
\frac{d}{dx}C_n^{(\alpha)}(x)- n(n+2\alpha)C_n^{(\alpha)}(x)= 0,
\eeq
with $\alpha=1$. The Legendre polynomial equation corresponds to
$\alpha=1/2$. All other hyperspherical coordinates will lead to
associated Geigenbauer equations with integer or half integer
$\alpha$.

The hyperspherical harmonics in four dimensions are given by:
\beq
Y_{nlm}(\theta,\phi,\psi)= A(n,l,m) \left\{
  \begin{array}{l}
      Y_{lm}(\theta,\phi) \;\;  U_n{}^l(\cos(\psi))~~~~~~~~~  0\leq l\leq n  \\
      Y_{|l|m}(\theta,\phi)\;\;  U_n{}^{l-1}(\cos(\psi))~~~   -n\leq l\leq -1
  \end{array}.
 \right.
\eeq
where $Y_{lm}(\theta,\phi)$ are the usual 3D spherical harmonics and
$U_n{}^l$ are associated Chebyshev type II functions defined by:
\bea
U_n{}^l(x)&=&(1-x^2)^{l/2} \frac{d^l}{dx^l} U_n(x) \nonumber \\
&=&(1-x^2)^{l/2} \frac{(-1)^n(n+1) \sqrt{\pi}}{2^{n+1} (n+1/2)!} \frac{d^l}
{dx^l}
\left[
 (1-x^2)^{-1/2} \frac{d^n}{dx^n} \left\{ (1-x^2)^{n+1/2} \right\} \right],
\eea
for $l\geq 0$ and by:
\beq
U_n{}^l(x)= \frac{(-1)^n(n+1) \sqrt{\pi}}{2^{n+1} (n+1/2)!} (1-x^2)^{l/2} 
\frac{d^{n+l+1}}{dx^{n+l+1}} \left[ (1-x^2)^{n+1/2} \right],
\eeq
for $l\leq -2$.
These hyperspherical harmonics form a complete orthogonal basis for the functions
of the angular variables in four dimensions.
\beq
\sum_{n=0}^\infty \sum_{l=-n}^n \sum_{m=-l}^l Y^*_{nlm}(r,\theta,\phi,\psi) 
Y_{nlm}(r^\prime,\theta^\prime,\phi^\prime,\psi^\prime)=
\frac{\delta(\psi-\psi^\prime)}{\sin^2 \psi} 
\frac{\delta(\theta-\theta^\prime)}
{\sin \theta} \delta(\phi-\phi^\prime).
\eeq

\beq
\int_{0}^\infty \int_{0}^{\pi} \int_{0}^{\pi} \int_{0}^{2\pi}
Y^*_{nlm}(r,\theta,\phi,\psi) Y_{n^\prime l^\prime m^\prime}(r,\theta,\phi,\psi)=\delta_{nn^\prime}
\delta_{ll^\prime}\delta_{ll^\prime}.
\eeq

\end{document}